# Lattice, Charge and Spin phase inhomogeneity in complex striped quantum matter

Antonio Bianconi[1,2,3] * Alessandro Ricci[4,1]

[1] *RICMASS Rome International Center for Materials Science Superstripes, via dei Sabelli 119A, 00185 Roma, Italy*
[2] *Institute of Crystallography, CNR, via Salaria Km 29.300, Monterotondo, Roma, 00015, Italy*
[3] *National Research Nuclear University, MEPhI, Kashirskoe sh. 31, 115409, Moscow, Russia*
[4] *Deutsches Elektronen-Synchrotron DESY, Notkestrasse 85, D-22607 Hamburg*
*email: antonio.bianconi@ricmass.eu*

**Abstract** The Superstripes 2016 conference, held on June 23-29, 2016 in the island of Ischia in Italy celebrated the 20th anniversary of this series of conferences. For 20 years structural, electronic, and magnetic phase inhomogeneities in quantum matter have been the scientific focus for a growing physics community interested in complexity in quantum matter. It has been the meeting point for different scientific communities facing the challenging project to unveil the complex space and time landscapes in quantum matter. The interesting spatial inhomogeneity length scale of multiple coexisting phase ranges from atomic to mesoscopic and the time fluctuations are spread over multiple time scales. The response of these materials changes using different experimental techniques with different spatial and time resolution probing different aspects of the quantum complexity.

After the discovery of high temperature superconductivity Alex Müller organized two important workshops [1,2] on phase separation and Jahn-Teller polarons in cuprates superconductors bringing together scientists active in these fields.

After ten years of research on the physics of cuprate perovskites, together with Alex Muller, we decided to organize the first conference on "Stripes and high temperature superconductivity" which was held in Rome in December 1996 [3]. We invited the most active research groups reporting evidence for strong magnetic interactions resulting in the formation of spin density wave, SDW, in the form of antiferromagnetic stripes intercalated by charge rich stripes observed by neutron diffraction experiments using neutrons emitted by nuclear reactors [5]. These results showed that doped holes get self organized and segregate in stripes in spite of being uniformly distributed. Clearly these results imply an unconventional strongly correlated metallic phase with breakdown of the rigid band model [6-8].

At the same time other researchers investigated copper perovskites developing the new x-ray spectroscopy methods [8-10] using x-ray emitted by large electron storage rings. XANES (x-ray absorption near edge structure) experiments provided evidence for itinerant doped holes in the oxygen 2p orbital symmetry [9]. The EXAFS (extended x-ray absorption fine structure) experiments reported the evidence of charge density wave, CDW, [10] with associated periodic lattice distortions or orbital density wave giving the formation of nanoscale lattice stripes and nanoscale inhomogeneity [11] in agreement with PDF analysis of neutron scattering data [12] and transport data [13,14]. These data were interpreted in terms of polaronic (of pseudo Jahn-Teller type) charge density wave [15] and in terms of the formation of textures of quantum stripes giving multigap superconductivity tuned at a Lifshitz electronic topological transition [16,17].

In these last 20 years lattice, magnetic and electronic complexity in the new high temperature superconductors, like magnesium diborides, iron based superconductors, pressurized sulfur hydrides, has been object of discussion at the Superstripes conferences. Evidence has been accumulated that high temperature superconductivity emerges tuning the chemical potential at a Lifshitz electronic topological transition in multiband electronic systems. A superstripes scenario has been proposed where the short range stripe order form nanoscale puddles [18]. The development of highly focused x-ray photon beams at novel synchrotron radiation facilities has allowed the resolution of the multiscale phase separation from atomic, to nanoscale, up to mesoscale using scanning x-ray diffraction [19-21].

The new emerging scenario shows a multiscale complexity where different nanoscopic orders coexist at the same time. In such landscape the spin-density-wave (SDW), charge-density-wave (CDW) and defects get organized forming spatially anticorrelated complex networks of nano puddles. This scenario is schematically represented by the Figure 1 where blue, red and gray-striped puddles refer to SDW, CDW and defects-rich orders, respectively.

For the 20th anniversary of the series of the Superstripes conferences and high temperature superconductivity more than 230 selected top level scientists get together in the island of Ischia in Italy for the Superstripes 2016 conference on June 23-29, 2016 [22].

The key research activity in the field of high temperature superconductivity remains today addressed to the synthesis of new complex multi-components materials formed by hetero-structures at atomic limits made of modules in the form of atomic layers, or atomic wires intercalated by other portions playing the role of spacer components. [23]

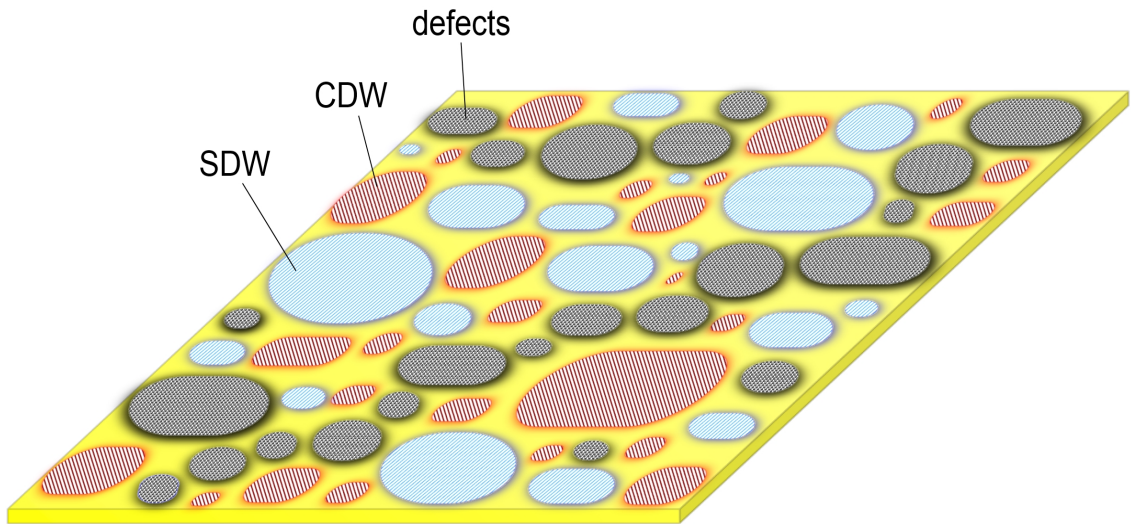

**Figure 1.** The complex *superstripes* scenario in doped strongly correlated electronic matter where at least three phases coexist in different spatial domains: spin-density-wave (SDW), charge-density-wave (CDW) and defects rich domains, where dopants get organized. These three complex networks form a very complex landscape of spatially anticorrelated nano-puddles where inhomogeneity ranges from atomic scale to nanoscale to mesoscale.

The electronic structure of these materials is manipulated by chemical doping or pressure to drive the system to unconventional insulator-to-metal transitions, like in cuprates, or to Lifshitz electronic topological transitions for the appearing of a new Fermi surface or for dimensionality changes like 3D to 2D dimensional transitions given by opening a neck in a tubular Fermi surface.

In the field of cuprates evidence for percolation of nanoscale puddles of the dopant oxygen chains in $YBa_2Cu_3O_{6.33}$ has been observed at the insulator to metal transition [24].

The multiple electronic components in cuprate superconductors has been confirmed by new results using nuclear magnetic resonance [25], charge order has been correlated with the presence of pseudogap electronic matter by high resolution photoemission [26]. Using magneto-transport experiment on $La_{1.48}Nd_{0.4}Sr_{0.12}CuO_4$, near the structural phase transition for the spacer layers from the rocksalt to fluorite structure at a critocal value of misfit strain avalanches have been observed as a function of temperature at the LTO-LTT structural transition confirming the lattice complexity [27].

In the field of the local atomic structure in iron based superconductors the double-well for the Fe-As bond fluctuations [28] and related lattice instabilities have been detected by EXAFS [29]. Scaling of the stress and temperature dependence of the optical anisotropy in Ba(Fe$_{1-x}$Co$_x$)$_2$As$_2$, the absence of long-wavelength nematic fluctuations in LiFeAs [30] and the vortex structure [31] and anisotropic superconducting gaps in Ba[Fe(Ni)]$_2$As$_2$. [32] have been observed and the europium valence state in related compound EuCo$_2$As$_2$ [33] has been investigated.

Carrier relaxation dynamics in organic superconductors has been studied by pump- and probe-polarization ultrafast spectroscopy [34,35]. Organic superconductors show similar electronic complexity as unconventional oxide superconductors showing the interplay between conducting and magnetic systems [36]. Felner has reported a large amount of data showing that amorphous carbon shows superconductivity and unusual magnetic [37]. The inhomogeneous electronic state has been found also in Bi$_2$Te$_3$ doped with manganese [38].

Finally an experimental evidence for magnetic correlations at the metal-to-insulator transition in a 2D correlated electronic gas has been discussed by Pudalov [39].

The hot topic of the conference has been multigap superconductivity [40-43]. The shape resonance superconductivity in nanoscale superconducting units has been discussed by Perali et al. [40]. In this scenario a first condensate at the BCS-BEC crossover resonates with a second condensate in the BCS regime. The physics of multi-condensate superconductors is very rich [41.42]. Yanagisawa has discussed the Nambu-Goldstone-Leggett and Higgs modes in multi-condensate superconductors [43].

Holographic models [44] and strongly correlated electronic wave functions [45] have been discussed for disordered systems without translational invariance. The emergent domain structures at the charge order-to-superfluid transition for 2D hard-core bosons [46] and the nanoscale phase separation in ferroelectric materials [47] have been described by advanced theories. The quantized massive gauge fields and hole-induced spin glass mechanism in underdoped cuprates [48] have been object of discussion as well as the presence of a pseudogap in the fluctuating charge-density wave phase of cuprates [49]. The physics of strongly correlated electronic systems deserves high interest [50-55] as well as spin-orbital physics in hybrid oxides [56,57]. The complex vortex matter in a superconductor in the BCS-BEC regime has been discussed by Salasnich [58]. In complex quantum matter electronics in reduced dimensions and the physics of electronic matter at the interface for nanoscale devices has been object of a relevant part of the conference [59-61]. An interesting presentation of the scientific contribution of Tomonaga to the development of BCS theory has been presented by Palumbo et al [62] and Turner has present a new *ab initio* approach to the development of new high temperature superconducting materials [63].